\documentclass[letterpaper, 10 pt, conference]{ieeeconf}  

\usepackage{epsfig} 
\usepackage{mathptmx} 
\usepackage{times} 
\usepackage{amsmath} 
\usepackage{amssymb}  
\usepackage{subfig}

\title{\LARGE \bf
DAARIA: Driver Assistance by Augmented Reality for Intelligent Automotive
}

\author{Paul George, Indira Thouvenin, Vincent Fr\'{e}mont and V\'{e}ronique Cherfaoui
\thanks{The authors are with \textit{University of Technology of Compi\`{e}gne
(UTC)},\textit{CNRS Heudiasyc UMR }7253\textit{, }\textit{Contact} {\tt\small firstname.familyname@utc.fr}}
}

\begin{document}
\maketitle
\thispagestyle{empty}
\pagestyle{empty}

\begin{abstract}

Taking into account the drivers'state is a major challenge for designing new advanced driver assistance systems. In this paper we present a driver assistance system strongly coupled to the user. \textsc{Daaria} \footnote{The authors are with \textit{University of Technology of Compi\`{e}gne(UTC)}, \textit{CNRS Heudiasyc UMR }7253\textit{, }{\tt\small firstname.familyname@utc.fr}} for Driver Assistance by Augmented Reality for Intelligent Automotive is an augmented reality interface informed by a several sensors. The detection has two focus: one is the obstacles position and the quantification of their dangerousness. The other is the driver behavior. Via a suitable visualization metaphor the driver can at any time perceive the location of relevant hazards while keeping his eyes ont the road. First results show that our method could be applied to vehicle but also to aerospace, fluvial or sea navigation.

\end{abstract}

\section{Introduction}

Driving relies on key concerns such as security and mediated perception for the driver: driving accidents are the 10th leading cause of death worldwide, killing every year 1,2 million people. The automotive industry has turned to support driving systems such as ABS (Anti-lock braking) or ESP (Electronic-Stability Program). These systems attempt to correct driving errors when they are happening. An original method is proposed in this paper : a Mobile Augmented Reality and Interactive Driving Assistance which provides the user the possibility to avoid obstacles and to anticipate driving difficulties.  We designed a display module with Augmented Reality (AR) linking a module of obstacles perception and a module of observation of the driver's behavior. We present in part 2 a state of the art of visualization metaphors for driving. Then in part 3 we describe our approach, and part 4 provides the \textsc{daaria} system design and realization for obstacle detection.  The first results are given in part 5 and a conclusion is proposed in part 6.

\section{Review of existing metaphors for the driver}

The visualization metaphor is defined by Averbukh \cite{c1} as "\emph{a map establishing the correspondence between concepts and objects of the application domain under modeling and a system of some similarities and analogies. This map generates a set of views and a set of methods for communication with visual objects.}"
These metaphors are used to inform the user by reusing standard interface concepts. It is essential to avoid cognitive overload when learning to use the new system.
In this section we will extract from literature some existing visualization metaphors for car driving application.
\subsection{Metaphor for navigation and planning assistance}
The navigation assistance helps the driver to choose a direction while the planning assistance allows the user to reach its destination without prior knowledge on the road topology or on the followed path. We present in this section different metaphors related to the navigation and planning aid systems.

\begin{figure}[thpb]

\subfloat[][]{\label{fig1a}
            \includegraphics[width=0.45\columnwidth,keepaspectratio]{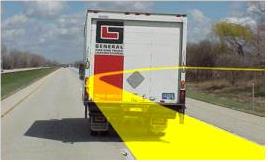}
            }
\subfloat[][]{\label{fig1b}
            \includegraphics[width=0.45\columnwidth,keepaspectratio]{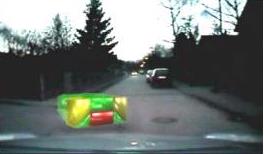}
                }
\qquad
\subfloat[][]{\label{fig1c}
            \includegraphics[width=0.45\columnwidth,keepaspectratio]{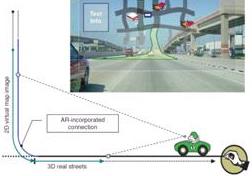}
                }
\subfloat[][]{\label{fig1d}
            \includegraphics[width=0.45\columnwidth,keepaspectratio]{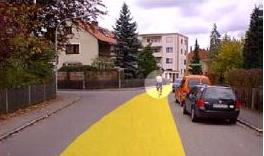}
                }
\qquad
\subfloat[][]{\label{fig1e}
            \includegraphics[width=0.45\columnwidth,keepaspectratio]{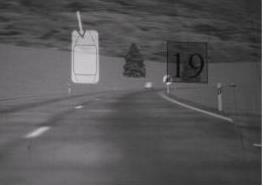}
                }
\subfloat[][]{\label{fig1f}
            \includegraphics[width=0.45\columnwidth,keepaspectratio]{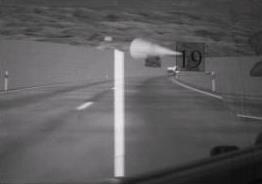}
                }
\caption{Metaphors for navigation, planning and driving assistance}
\end{figure}

Narzt et al. \cite{c5} \cite{c6} describe two metaphors types in the field of car assistance system.  The metaphor of the augmented road (Fig. \ref{fig1a}) provides the user a highlighted way to follow when looking directly at the road. For example, it allows to notice that a back exit has been missed or to find the right exit on a roundabout without having to check. This metaphor is for planning assistance even if the information presented to the user is a middle term one (usefull till the next intersection). Another metaphor present a virtual car to follow as it accelerates, brakes or activates turn signal. In this case, the information is a short term one (about one second) (cf Fig.\ref{fig1b}).

The metaphor of unrolled map \cite{c4} is an improvement of the metaphor of the augmented road. The aim is to give the user global knowledge of the environment. A map is held as and when the vehicle is moving. The part which is on the ground indicates the path to be taken immediately. The other part in the sky shows the structure of the surrounding road network. A curved area allows a continuous transition between the two modes of representation (Fig. \ref{fig1c}). It provides a knowledge of the road network in the medium term, using the technique of LOD (Level Of Detail).
The results of tests in simulation (i.e. virtual reality) show a reduction in navigation errors and problems caused by divided attention. This metaphor has been developed by CMU on the new HUD system "full windshield" of General Motors in order to verify these results in real situations.

\subsection{Metaphor for driving safety}
We define the driving safety assistance as all resources used to provide the user with information needed to drive as safely as possible. The aim is to overcome inattention and to compensate the ability to detect hazards.

The metaphor of the highlighter shown in Fig. \ref{fig1d} allows to highlight some of the details existing in the driver's field of view in order to draw his attention. It is possible to highlight other vehicles, pedestrians, and the lane limits\cite{c5}.
We can notice that the highlight of the road is similar to the augmented road metaphor. However, the context and the goals are not the same. Indeed, in the first case, the aim is to guide the driver along a route. In the second case,the aim is to help the user to locate the path when the lane is not clear or when the visibility is poor.

The metaphor of the radar \cite{c8} looks like a top view of the vehicle in two dimensions. Imminent hazards are marked with an arrow indicating their direction. The user can see the dangers that are on front of the vehicle (Fig. \ref{fig1e}). This metaphor is exocentric (i.e. external to), also the user must perform  mental transformations to bring the information in the egocentric reference. The exocentric metaphors are not suitable for driving assistance.

The metaphor of the vane \cite{c8} is presented as a three-dimensional arrow pointing to imminent danger (Fig. \ref{fig1f}). The arrow is attached to a virtual pole in front of the car, and helps preventing cognitive changes (otherwise some subjects perform a mental translation of the arrow to the location of their head).
Unlike the metaphor of the radar, the metaphor of the wind vane keeps an egocentric view while providing the driver with information on items that are not in his visual field. The advantage of staying in egocentric visualizations is that it frees the driver to perform mental transformations before processing information.

\subsection{Metaphors to discover points of interest}
\begin{figure}[thpb]

\subfloat[][]{\label{fig2a}
            \includegraphics[width=0.45\columnwidth,keepaspectratio]{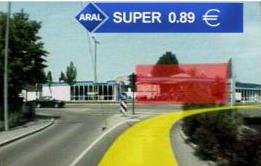}
            }
\subfloat[][]{\label{fig2b}
            \includegraphics[width=0.45\columnwidth,keepaspectratio]{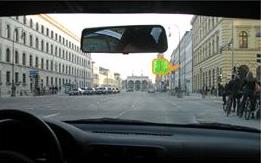}
                }
\caption{Metaphors to discover Points of interest}
\end{figure}

The discovery of points of interest is a mean to provide the user with additional knowledge without distraction.
Narzt \cite{c5} offers the annotation metaphor with contextual information. He suggests to take into account the vehicle's state in order to provide information suitable to the driver's needs such as: location of a gas station and the fuel price when missing (Fig.\ref{fig2a}).
A variant of the annotation metaphor is the use of a haptic touch pad representing the road and points of interest. The user can touch the points of interest to select them. He has both visual feedback via augmented reality, and haptic feedback through the pavement. He can finally click on a point of interest (always via the keypad), and get more information.
Adding increased interaction in this context leads to interact with the interface safely.

\subsection{Conclusion on metaphors uses for driving assistance}
Many metaphors are available in the context of car driving. Displayed all together they would lead to an overload of the driver's visual space. Our first objective will be to design a relevant metaphor.
In addition, with existing metaphors, authors often assume that a future technology will be able to display on the whole windshield. We assume we have not this possibility. Our second objective will be to develop a prototype in order to conduct experiments for our solution and to realize an adapted interface.

\section{ Proposed Approach}
Our objective is to integrate the visualization metaphor in the design process of driving assistance system. In order to propose a visualization metaphor adapted to the driving situation, it is  necessary to know what is the current situation and what the driver looks at. So we propose the coupling of an obstacle detection module with a system monitoring the driver in order to develop a display module for an application of Augmented Reality (AR). This section is organized as follows : we describe first the obstacle detection and the eye tracking system used for this study; then we describe our metaphor.

\subsection{Obstacle detection and eye tracking system.}

\begin{figure}[thpb]

\subfloat[][]{\label{fig3a}
            \includegraphics[width=0.55\columnwidth,keepaspectratio]{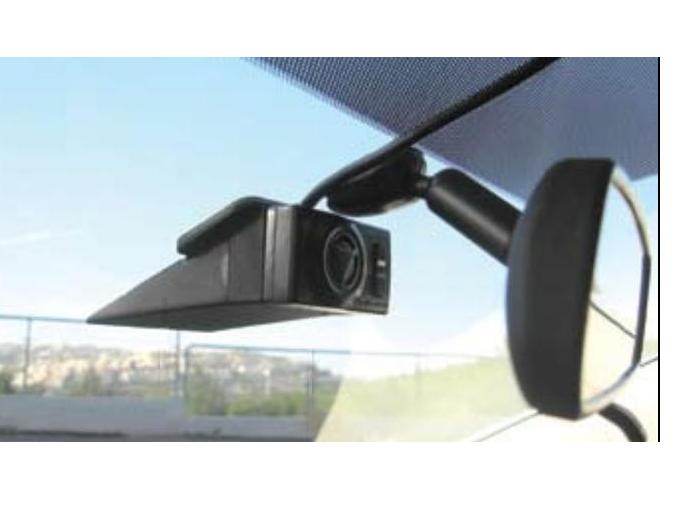}
            }
\subfloat[][]{\label{fig3b}
            \includegraphics[width=0.35\columnwidth,keepaspectratio]{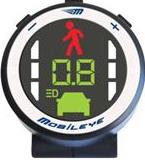}
                }
\caption{The system Mobileye and the public interface}
\end{figure}

To achieve obstacles detection, we will use the ADAS (Advanced Driver Assistance Systems and Driver Assistance) Mobileye$^\copyright$. A camera is mounted on the windshield of the vehicle (Fig.\ref{fig3a}) and a computer (chipset) retrieves the video stream and processes it to extract more useful information for driver assistance applications:  positions of pedestrians and vehicles ( Cars, trucks, bicycles and motorcycles), vehicle speed, road markings etc... The public version of Mobileye offers an interface for driver assistance notifying the presence of vehicles, pedestrians and lane (Fig.\ref{fig3b}). The system is also equipped with loudspeakers to broadcast warnings.
The professional version provides access to information calculated by the system and allows to exploit them differently from the original system. The data produced are broadcasted on the CAN bus of the vehicle. Video signal is not an output of the system.

\begin{figure}[thpb]
\subfloat[][]{\label{fig4a}
            \includegraphics[width=0.45\columnwidth,keepaspectratio]{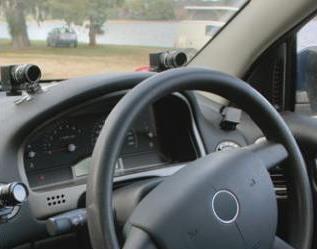}
            }
\subfloat[][]{\label{fig4b}
            \includegraphics[width=0.45\columnwidth,keepaspectratio]{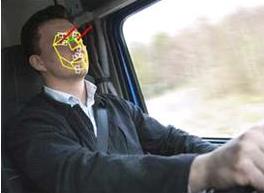}
                }
\caption{The eye-tracker Facelab}
\end{figure}

To adapt the display to the metaphor, we have chosen to integrate the driver's behavior to the system, especially his gaze. The idea is to inform the driver only when he needs help.
FaceLab$^\copyright$ is a system able to capture head and eyes position and orientation. Two infrared emitters illuminate the user's face, two cameras are located in the configuration of the stereo vision (cf Fig. \ref{fig4a}). A computer processes the images to extract the following information : the position and orientation of the head, the position and orientation of the eyes, opening of the eyelids, pupil size, the frequency of eye blinking as shown in fig. \ref{fig4b}).\\
FaceLab is a commercial product and outputs are saved in a file or sent in the network.

\subsection{The visualization metaphor}


In order to help the driver with obstacle detection, we have chosen the metaphor of the weathervane. This metaphor has two main advantages : it is egocentric and it can be used to indicate several dangers at once. The metaphor proposed for the restitution to the driver is defined as follow.


 \subsubsection{Indication of the type of danger}  Rather than using a color code to indicate the nature of the dangers, we will refer to symbols which are familiar to the driver: those of the highway code (Fig. \ref{fig5a}).
  \subsubsection{Indication of dangerousness}To highlight the most critical dangers, we have adopted an intuitive color-code : the gradient green / yellow / orange / red. Indeed, it is generally accepted that red and green are associated respectively danger and safety (Fig. \ref{fig5b}).
 \subsubsection{Indication of criticality} In order to solve the problem of overlapping arrows and prioritize the display of arrows indicating the most critical dangers, we decided to convey information on the criticality of the danger through the height of the arrows. They are sorted so that the more dangerous, the more visible, thus allowing the driver to process them in an optimal way (Fig. \ref{fig5c}).
  \subsubsection{ Animation of metaphor} 
\begin{figure}[thpb]
\qquad
\subfloat[][]{\label{fig5a}
            \includegraphics[width=0.45\columnwidth,keepaspectratio]{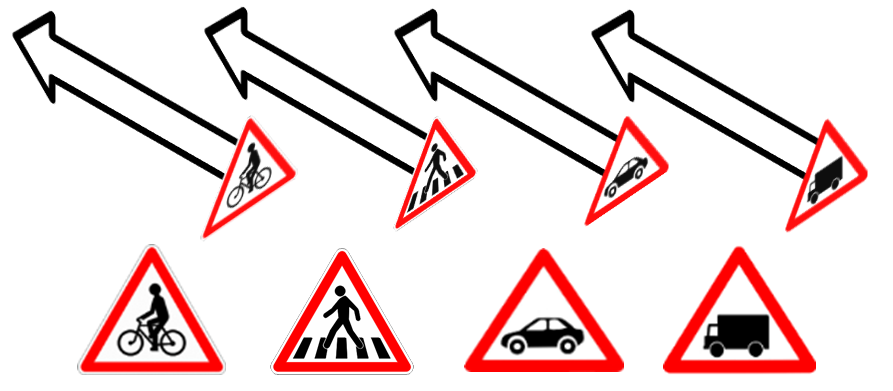}
            }
\subfloat[][]{\label{fig5b}
            \includegraphics[width=0.45\columnwidth,keepaspectratio]{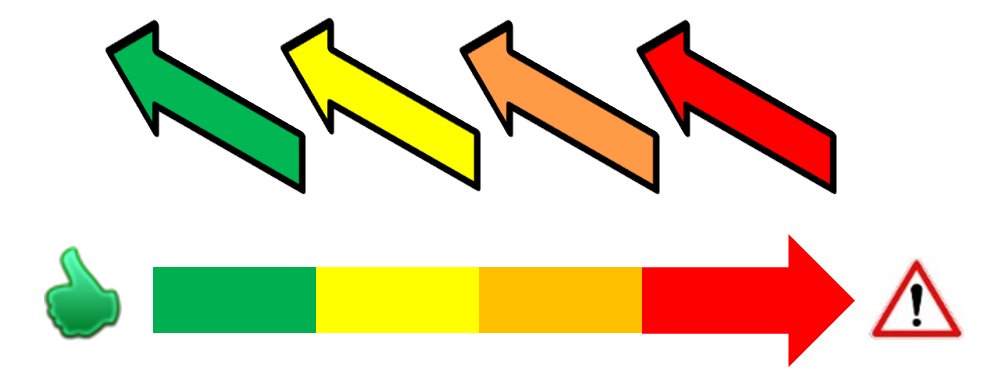}
                }
\qquad
\subfloat[][]{\label{fig5c}
            \includegraphics[width=0.40\columnwidth,keepaspectratio]{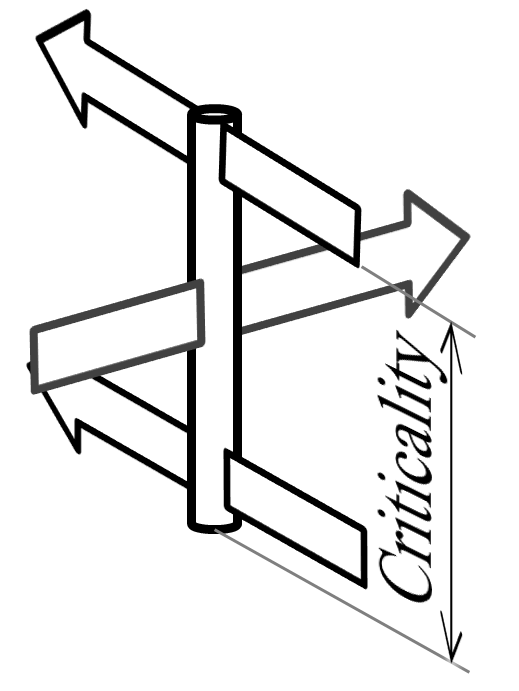}
                }
\subfloat[][]{\label{fig5d}
            \includegraphics[width=0.40\columnwidth,keepaspectratio]{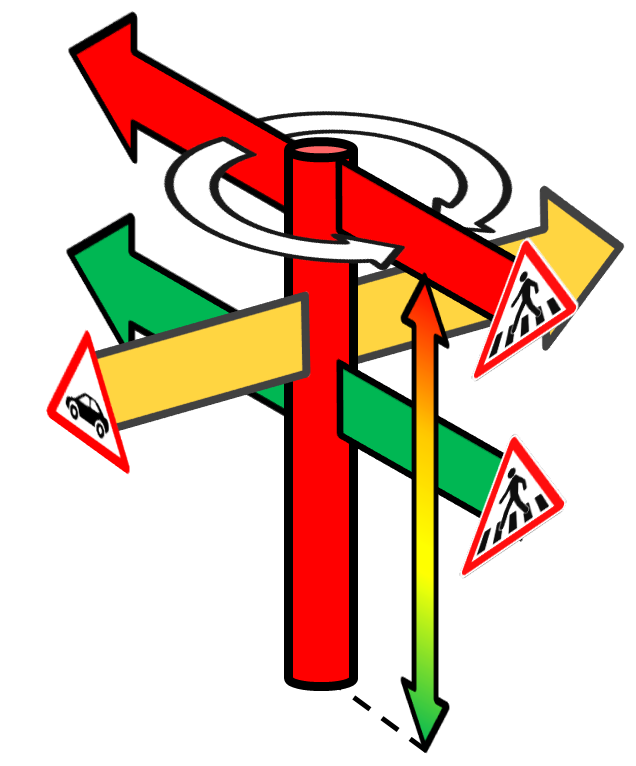}
                }
\caption{Metaphors for obstacle detection assistance }
\end{figure}

To make the metaphor more pleasing to the eye, and give it a more credible behavior, we decided to animate it. In practice, the arrows motions are physics-driven, hence smoothing variations of metaphor states. Fig. \ref{fig5d} is a diagram showing the design of our metaphor building on the original weathervane metaphor.

\section{\textsc{Daaria} system}

\subsection{System Design}

\begin{figure}[thpb]
      \centering
      \includegraphics[width=0.9\columnwidth]{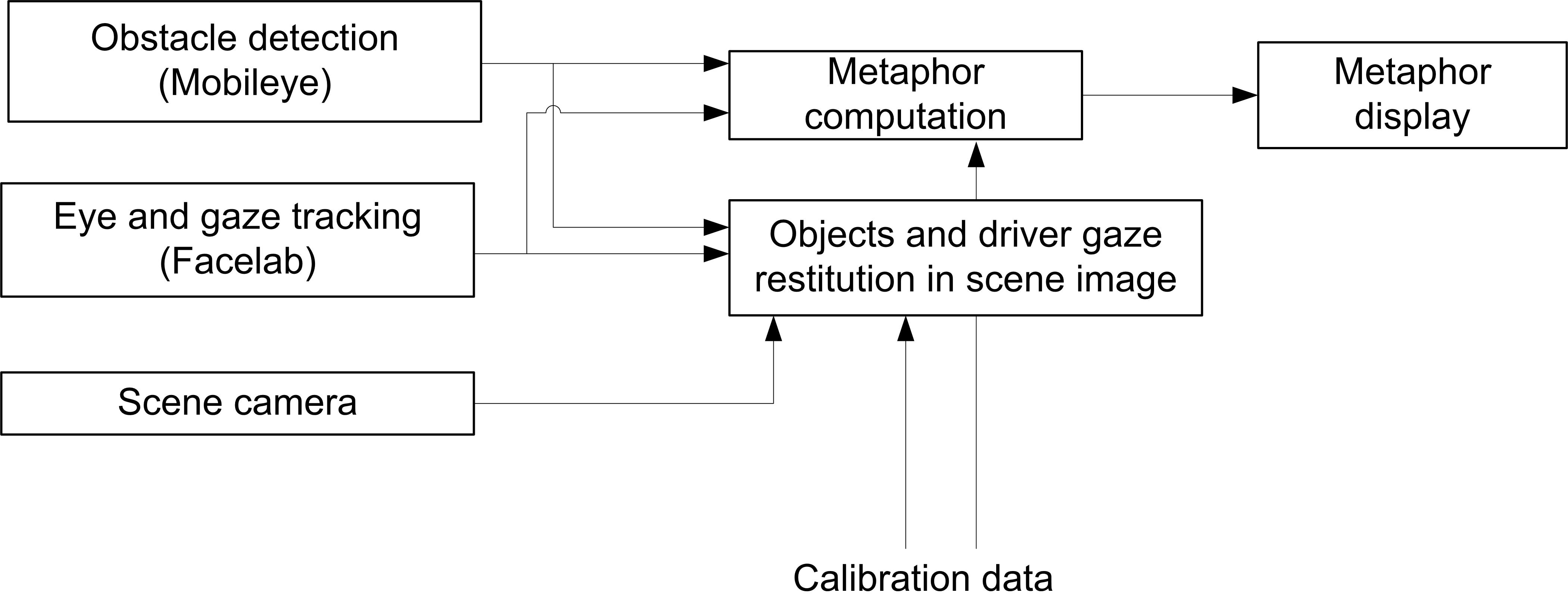}
      \caption{\textsc{Daaria} Architecture}
      \label{fig17}
   \end{figure}

 The \textsc{Daaria} system (Driver Assistance by Augmented Reality for Intelligent Automotive) is an augmented reality interface informed by two detection module. One is the obstacles position and the quantification of their dangerousness. The other is the user behavior. The  \textsc{Daaria} architecture is described in figure \ref{fig17}.

\begin{figure}[thpb]
      \centering
      \includegraphics[width=0.8\columnwidth]{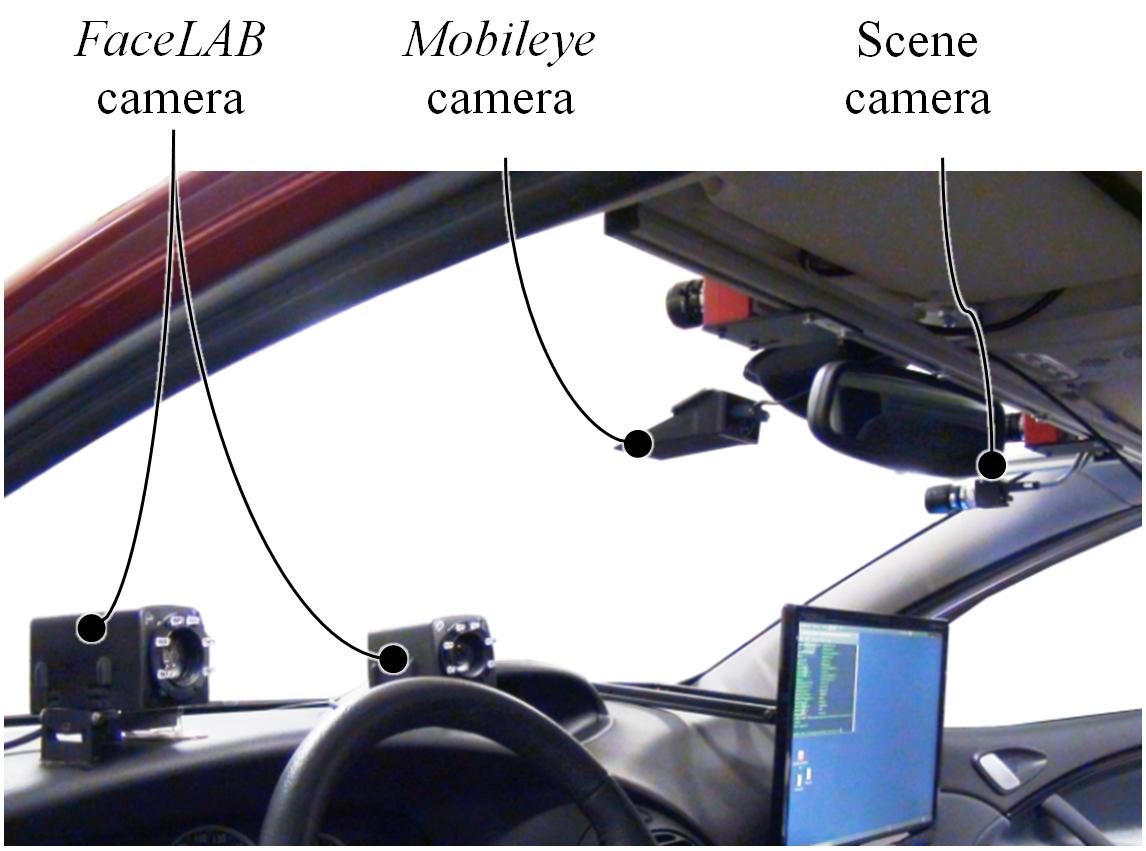}
      \caption{Integration on board CARMEN}
      \label{fig18}
   \end{figure}

The experimental platform \textsc{carmen}  has been used to develop \textsc{Daaria}. \textsc{Carmen} is an experimental vehicle dedicated to research on intelligent vehicles. The Figure \ref{fig18} shows the cameras of Mobileye and FaceLab systems. Another camera (scene) has been added in order to have a video output of the scene and to monitor the system. In the same way, the display allows to check that all components are working correctly when experimenting.

\subsection{Calibration method}

\begin{figure}[thpb]
      \centering
      \includegraphics[width=0.8\columnwidth]{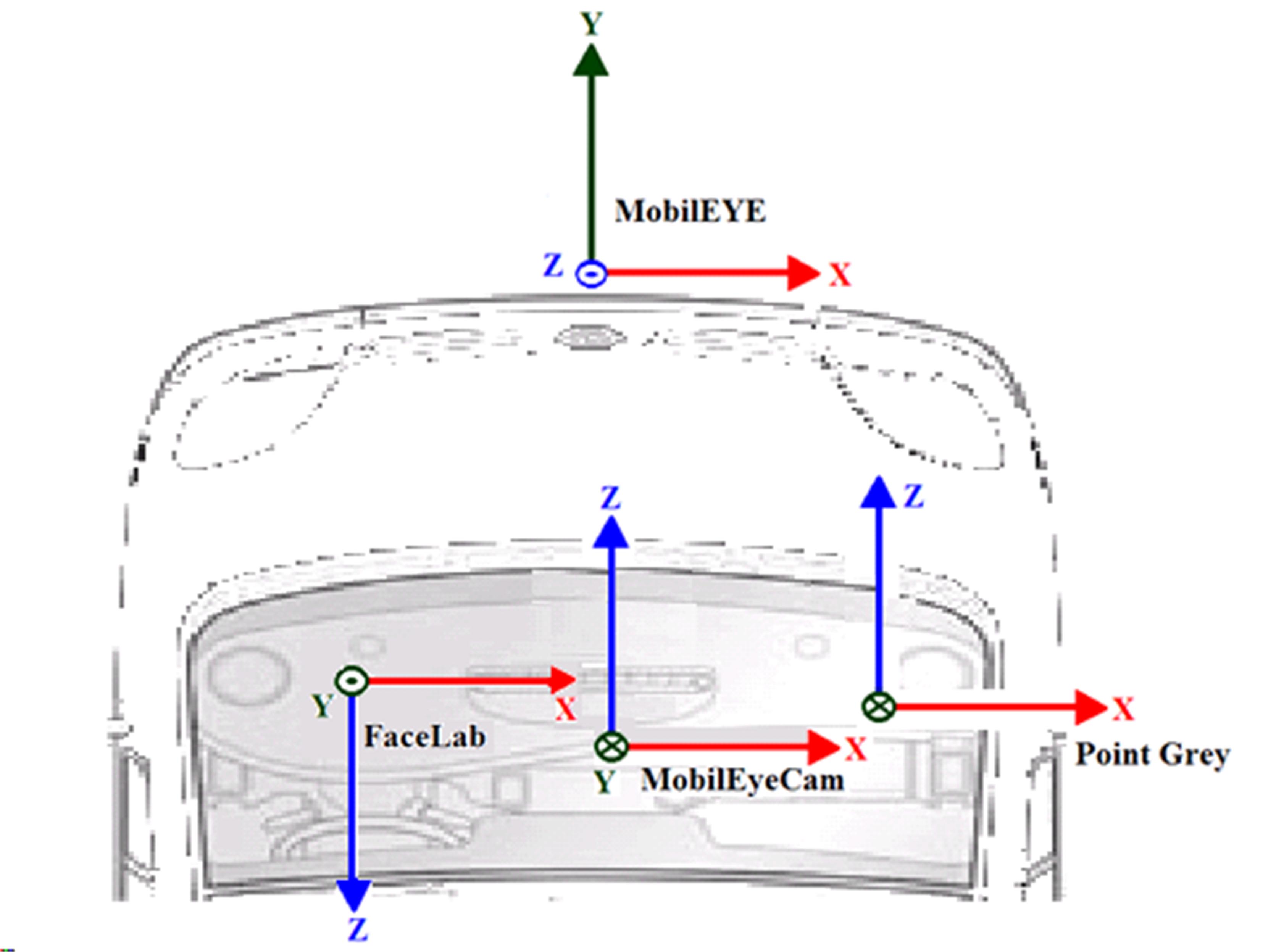}
      \caption{Frames}
      \label{fig19a}
   \end{figure}

The data to combine in \textsc{Daaria} come from heterogenous sources : Mobileye, FaceLAB and a wide angle camera. Since the geometrical informations are expressed in different frames (see Fig. \ref{fig19a}) , it is crucial to perform an extrinsic calibration procedure between the systems \cite{c10}. For each camera system, we use the Caltech Calibration toolbox \cite{c12} to estimate the intrinsic parameters of each camera. If the cameras have enough overlapping, the calibration toolbox can be used to estimate the extrinsic parameters i.e. rotation and translation between the cameras frames. In our case, the Mobileye and the FaceLAB cameras are looking at opposite direction, therefore classical calibration procedure is impossible. Moreover, it is impossible to access to the video images in the FaceLAB system. The only available information comes from the gaze tracker which gives the eyes direction in the FaceLAB frame.

\begin{figure}[thpb]
      \centering
      \includegraphics[width=0.8\columnwidth]{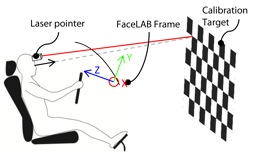}
      \caption{Calibration System}
      \label{fig19b}
   \end{figure}

  Therefore we developed a new idea to solve the calibration problem (see Fig. \ref{fig19b} ):
  \begin{itemize}
  \item We take pictures of the calibration target with the Mobileye System.
  \item During the pictures acquisition, someone seats on the driver side, looking at some reference point on the target and we save the gaze direction vector.
  \item Simultaneously, we place a Laser pointer on the driver's forehead pointing on the calibration target, and we save the measured distance.
  \item If one repeats the procedure several times, one obtains 3D points in the Mobileye frame and in the FaceLAB frame. Then we use a classical registration approach such as ICP \cite{c11} to estimate the transformation between the two frames.
  \end{itemize}

\subsection{Objects and driver gaze restitution in scene image}
\begin{figure}[thpb]
\subfloat[][]{\label{fig20a}
            \includegraphics[width=0.20\columnwidth,keepaspectratio]{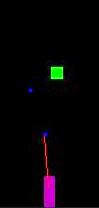}
            }
\subfloat[][]{\label{fig20b}
            \includegraphics[width=0.75\columnwidth,keepaspectratio]{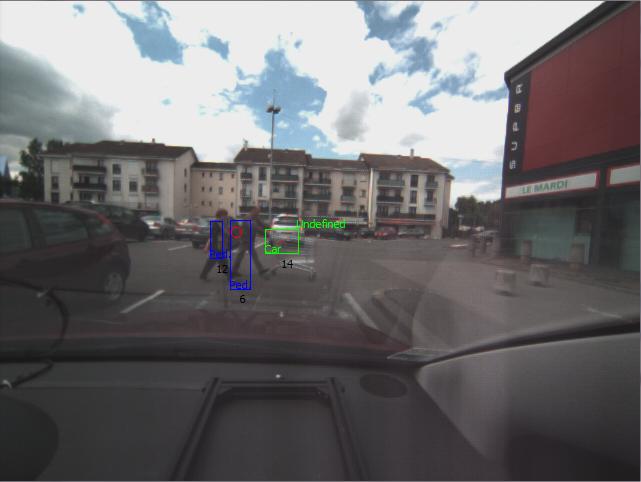}
                }
\caption{Objects and Driver gaze Restitution}
\end{figure}

This module is developed in order to have the possibility to check and monitor the metaphor computation. It calculates the projection of the driver's gaze direction and detected objects on two views : a bird view (Fig \ref{fig20a}) and a scene view given by the scene camera (Fig. \ref{fig20b}). This is done thanks to calibration data. The blue boxes represent pedestrians, the green boxes represent the cars and red line or circle gives the direction of the driver gaze.

\subsection{Metaphor computation}
The metaphor computation module calculates the configuration to be taken by the metaphor of the weathervane. To achieve this, we combine information from the viewing direction (FaceLAB) and the obstacles location (Mobileye) and can get obstacles information looked at. Determining obstacles dangerousness relies on several criteria:
\begin{itemize}
  \item Object dynamics : the stationary objects are not considered for the metaphor.
  \item Driver's visual activity: it avoids  to point out a danger checked by the driver. The current implementation is very simple : it carries out tests of intersection between the eye direction and obstacles.
  \item Time-to-Collision (TTC) : using the status of ego-vehicle and information about the detected objet, we can approximate a TTC. The lower, the more dangerous. The arrows in the weathervane metaphor are ordered by level of danger.
\end{itemize}

\subsection{Display device}

\begin{figure}[thpb]
\subfloat[][]{\label{fig21a}
            \includegraphics[width=0.48\columnwidth,keepaspectratio]{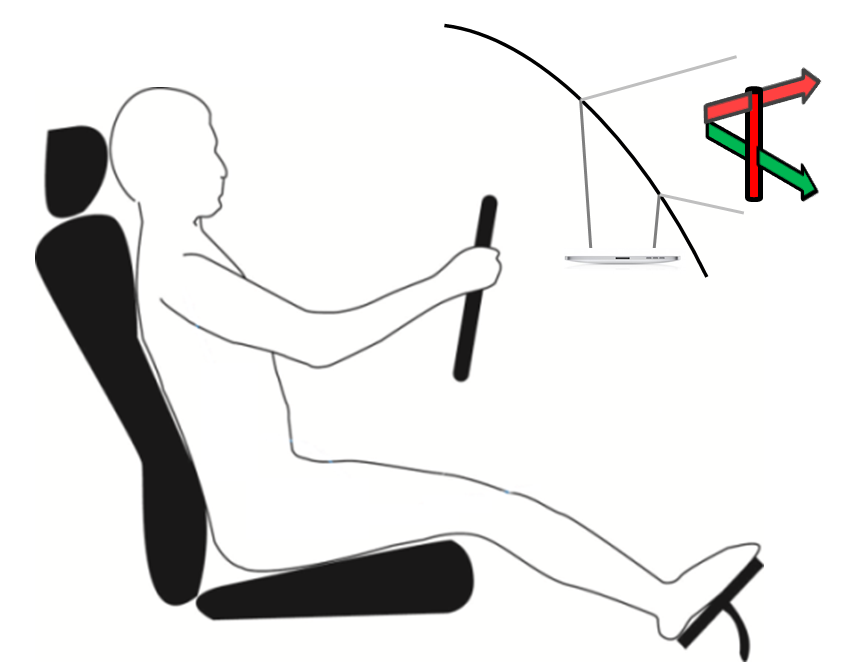}
            }
\subfloat[][]{\label{fig21b}
                 \includegraphics[width=0.42 \columnwidth]{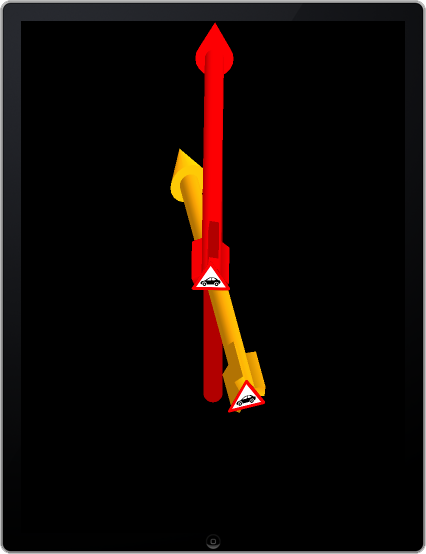}
            }
\caption{Tablet as HUD, displaying the metaphor of the weathervane}
\label{fig21}
\end{figure}

The most suitable device for an AR interface compatible with the the driving task would be the HUD (Head Up Display). Its main advantage over the HDD (Head Down Display) is the driver time saving. Indeed, the eyes back and forth between the road and the dashboard deprive the driver of valuable time, as the need to constantly refocus his eyes to distance changes.
HUD solves both problems: it is close to the road, which limits the round-trip of gaze and in addition the image formed is located behind the windshield, which reduces the visual rehabilitation. We proposed a low-cost solution that uses a tablet-PC as a HUD. The solution is to put it under the windshield of the car and to display our metaphor as shown in Fig. \ref{fig21}.

\section{ Preliminary results and discussions}
\begin{figure*}[t]
\begin{centering}
\includegraphics[width=0.75\textwidth]{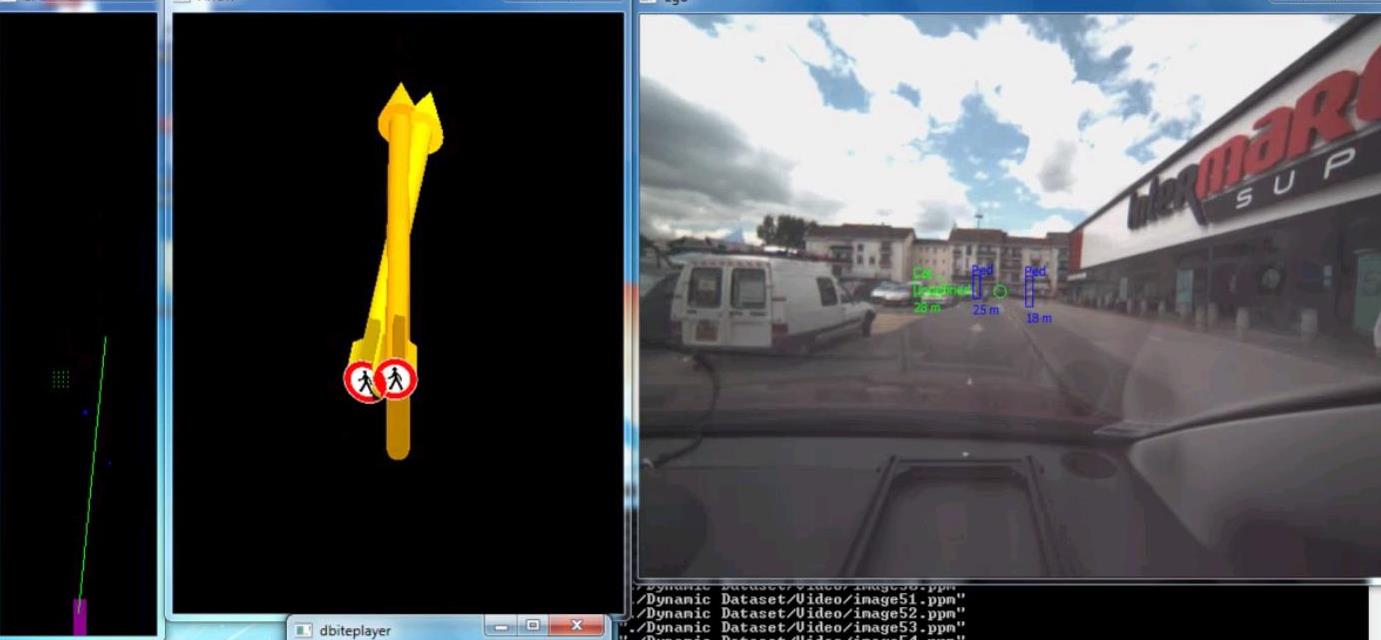}
\par\end{centering}

\caption{Replay for result analysis\label{fig22}}
\end{figure*}

The first real road tests allowed us to validate the choice of  components as well as the feasibility of our approach. \textsc{Carmen} is able to collect data that we can replay for analysis (Figure \ref{fig22}).

\subsection{Characterization of the consideration of an obstacle} To integrate the driver's behavior in our system, we chose a very simple rule based on the intersection of vector of the gaze of the driver and detected obstacles. This simple rule has several basic flaws: no consideration of gaze tracking, the field of view is not a vector, the imprecise or incorrect detection  can lead to wrong conclusions.
We could propose several improvements concerning the visual attention model, the finer characterization of the obstacle consideration and the cognitive map managing of the driver taking into account the information life cycle.
  \subsection{Characterization of dangerous obstacles} To integrate the situation analysis to DAARIA, we chose a simple rule: the closer the obstacle, the more dangerous. More information could be taken into account: the obstacle nature (pedestrian or vehicle), its trajectory and its speed, its behavior (a pedestrian about to cross the street), the consideration of traffic (for example, priority of the drivers on the right).

\section{Conclusion and outlooks}
A prototype of driver assistance system in augmented reality has been presented. It is based on the weathervane metaphor adapted for a use in real conditions. The metaphor is computed according to the driving situation and the driver's attention.
The driver can perceive at any time the location of the dangers without leaving the eyes of the road. The system adapts its behavior to provide him  the most relevant information.
A lot of problem of integration (calibration, synchronization - not described here- and data exchange) have been resolved to conduct first experiments and prove the feasibility.

Future works concern three major directions. The first one is the uncertainties consideration propagated in this complex system and especially the uncertainties of detected objects \cite{c2} and of the driver's attention. The second axe for future research is focused on the metaphor improvement. The third axe is the analysis of metaphor performances in virtual environnement \cite{c3} in order to validate it and test it with critical scenarios.

\section{ACKNOWLEDGMENTS}
The authors gratefully acknowledge the contribution of G\'erald Dherbomez.


\begin{thebibliography}{99}


\bibitem{c11}
K. Arun, T. Huang and S. Blostein, ?Least-squares Fitting of Two 3-D Point sets,? {\it IEEE Trans. Pattern Anal. Mach. Intell.},Vol. 9, No. 5, pp. 698-700; 1987.

\bibitem{c1}
Averbukh, V., M. Bakhterev, et al., Interface and visualization metaphors. {\it Proceedings of the 12th international conference on Human-computer interaction: interaction platforms and techniques}, Beijing, China, Springer-Verlag,2007, pp. 13-22.


\bibitem{c2}
Fayad, F. and Cherfaoui, V. and Derbhomez, G.,Updating confidence indicators in a multi-sensor pedestrian tracking system, {\it Proc. IEEE Intelligent Vehicles Symposium}, 2008, Eindhoven.


\bibitem{c3}
Fricoteaux, L. and Mouttapa Thouvenin, I. and Olive, J., Heterogeneous Data Fusion for an Adaptive Training, {\it Informed Virtual Environment. Systems (VECIMS)}. Ottawa, Canada, 2011.

\bibitem{c4}
Kim, S. and A. K. Dey (2009). Simulated augmented reality windshield display as a cognitive mapping aid for elder driver navigation. Proceedings of the 27th international conference on Human factors in computing systems. Boston, MA, USA, ACM, 2009, pp. 133-142.

\bibitem{c5}
Narzt, W., G. Pomberger, et al., A new visualization concept for navigation systems, Berlin, ALLEMAGNE, Springer, 2004.

\bibitem{c6}
Narzt, W., G. Pomberger, et al., Pervasive information acquisition for mobile AR-navigation systems, 2003.


\bibitem{c10}
S. A. Rodriguez F., V. Fremont and P. Bonnifait. Extrinsic Calibration between a Multi-Layer Lidar and a Camera. {\it In proceedings of IEEE Multisensor Fusion and Integration for Intelligent Systems}, August 20-22, Korea University, Seoul, Korea, 2008.

\bibitem{c7}
Spies, R., M. Ablameier, et al. ,Augmented Interaction and Visualization in the Automotive Domain, {\it Proceedings of the 13th International Conference on Human-Computer Interaction. Part III: Ubiquitous and Intelligent Interaction} San Diego, CA, Springer-Verlag, 2009, 211-220.

\bibitem{c8}
Tonnis, M. and G. Klinker, Effective control of a car driver's attention for visual and acoustic guidance towards the direction of imminent dangers , {\it Proceedings of the 5th IEEE and ACM International Symposium on Mixed and Augmented Reality}, 2006, pp. 13-22.

\bibitem{c9}
Tonnis, M., C. Sandor, et al., Experimental Evaluation of an Augmented Reality Visualization for Directing a Car Driver's Attention, {\it Proc. of the 4th IEEE/ACM International Symposium on Mixed and Augmented Reality},2005,pp. 56-59.



\bibitem{c12}
http://www.vision.caltech.edu/bouguetj/

\end{thebibliography}
\end{document}